# Symmetrizing cathode-anode response to speed up charging of nanoporous supercapacitors

Tangming Mo,[1,2] Liang Zeng,[1,2] Zhenxiang Wang,[1,2] Svyatoslav Kondrat,[3,4,5] Guang Feng[1,2,*]

[1] State Key Laboratory of Coal Combustion, School of Energy and Power Engineering, Huazhong University of Science and Technology, Wuhan 430074, China

[2] Nano Interface Centre for Energy, School of Energy and Power Engineering, Huazhong University of Science and Technology, 430074, China

[3] Institute of Physical Chemistry, Polish Academy of Sciences, Kasprzaka 44/52, 01-224 Warsaw, Poland

[4] Max-Planck-Institut für Intelligente Systeme, D-70569 Stuttgart, Germany

[5] IV. Institut für Theoretische Physik, Universität Stuttgart, D-70569 Stuttgart, Germany

*Corresponding author's email: gfeng@hust.edu.cn

**Abstract**

Asymmetric behaviors of capacitance and charging dynamics in the cathode and anode are general for nanoporous supercapacitors. Understanding this behavior is essential for the optimal design of supercapacitors. Herein, we perform constant-potential molecular dynamics simulations to reveal asymmetric features of porous supercapacitors and their effects on capacitance and charging dynamics. Our simulations show that, counterintuitively, charging dynamics can be fast in pores providing slow ion diffusion and vice versa. Unlike electrodes with single-size pores, multi-pore electrodes show overcharging and accelerated co-ion desorption, which can be attributed to the subtle interplay between the dynamics and charging mechanisms. We find that capacitance and charging dynamics correlate with how the ions respond to an applied cell voltage in the cathode and anode. We demonstrate that symmetrizing this response can help boost power density, which may find practical applications in supercapacitor optimization.

**Keywords**: nanoporous carbon; charging dynamics; charge storage mechanism; overfilling; overcharging

# Introduction

Supercapacitors have attracted significant attention owing to their fast charging/discharging rate and long cycle life.[1-3] However, their wide application is limited by their moderate energy density compared with batteries.[4-6] To address this issue, porous carbons have been used as electrodes in supercapacitors due to their high specific surface area and large specific capacitance.[7-11] Among capacitive behaviors of porous carbon supercapacitors, an exciting feature that the cathode and anode have unequal capacitance has been reported in both experiment and modeling works.[12-17] Electrochemical measurements on supercapacitors composed of carbide-derived carbons (CDCs) and an organic electrolyte of tetraethyl-phosphonium tetrafluoroborate in acetonitrile ([TEA][$BF_4$]/ACN) revealed that the capacitance of the negatively polarized electrode is much larger than the positive one,[12] while supercapacitors with CDCs and an ionic liquid (IL) of 1-ethyl-3-methylimidazolium bis(trifluoromethanesulfonyl)imide ([EMIM][TFSI]) exhibit a higher capacitance in the positive electrode.[13] Experiments on CDCs with [EMIM][TFSI] and [EMIM][$BF_4$] ILs disclosed that such an asymmetric behavior of the capacitance in negative and positive electrodes limits their voltage window and thereby the energy density.[16] Recently, supercapacitors, involving the positive electrode with microporous carbons for the exclusive electrosorption of small anions and the negative electrode of mesoporous carbons accessible to both ions, have been found to provide high energy storage capacity.[18] These studies indicate that the ion and pore size significantly impact the asymmetric capacitance behavior, which affects their energy storage.

Dynamic features of electrolyte ions in porous electrodes have also been revealed to exhibit an asymmetric behavior at cathode and anode.[2, 19-24] Tsai *et al.* utilized electrochemical quartz crystal microbalance (EQCM) to explore ion dynamics during the charging process in CDCs with [EMIM][TFSI] and found that the charge storage is dominated by the counterion adsorption under negative and highly positive polarization, and by ion-exchange under lower positive polarization.[20] Pean *et al.* performed molecular dynamics (MD) simulations of CDC electrodes in 1-butyl-3-methylimidazolium hexafluorophosphate ([BMIM][PF$_6$]) and predicted that cations are less mobile than anions during the charging process.[25] The diffusion of ions inside nanopores was found to correlate with the response of cations and anions inside the pores to a cell voltage.[26-27] Using NMR, Forse et al. found that the strongly reduced diffusion of ions in the negative electrode and only slightly varying diffusion in the positive electrode are associated with cation adsorption under negative polarization and ion-exchange under positive polarization, respectively.[26] However, it remains unclear what the origin of this asymmetric behavior is and how it affects the charging processes.

In this work, we perform MD simulations of symmetric and asymmetric supercapacitors to investigate asymmetric features of the cathode and anode response and their relation to capacitance and charging dynamics. Our simulations reveal that the capacitance and charging dynamics in both symmetric and asymmetric supercapacitors correlate with the ion response in the cathode and anode; symmetrizing this response can help speed up the charging process. Surprisingly, we find that despite fast diffusion of in-pore ions, narrow pores can exhibit slow charging dynamics caused by overfilling during the initial stage of charging, leading to slow co-ion desorption. However, this slow co-ion desorption can be accelerated in electrodes with multi-pore.

## Simulation methods and models

### Molecular dynamics simulation

We study the charging of supercapacitors with two electrodes, each containing one or two slit-shaped pores (**Fig. 1**). In these systems, an electrode that adsorbs cations in response to an applied potential will be interchangeably called a negative electrode or a cathode. Likewise, an electrode adsorbing anions will be called an anode or a positive electrode. In all simulations, we use a positive cell voltage between the cathode and anode. Systems with single pore electrodes, illustrated in **Fig. 1a**, are denoted as $d_{s1}//d_{s2}$ where s1 and s2 are the pore sizes of the positive and negative electrode, respectively. Systems with multiple pore sizes, shown in **Fig. 1b**, are named $d_{s1}d_{s2}//d_{s3}d_{s4}$, which means that a negatively charged electrode has two pores with sizes s1 and s2 and the positively charged electrode has two pores with sizes s3 and s4. In all simulations, the pore length was 8 nm. The simulation box sizes were $3 \times 3.53 \times 32$ nm$^3$ for single-pore and $3 \times 7.06 \times 32$ nm$^3$ for multi-pore systems. Periodic boundary conditions were applied in all three directions.

We used the four-site coarse-grained model of [EMIM][BF$_4$],[28] containing a large elongated cation and a smaller spherical anion (**Fig. S1**), and the Lennard-Jones model for carbon atoms.[29] Based on the [EMIM][BF$_4$] model, the average ion size is about 0.5 nm. To study ion-size--pore-size relation, we focused on three pore widths: 0.5 nm (comparable to the ion size), 0.45 nm (ionophobic pore[30]), and 0.75 nm (1.5 times the ion size). The electrolyte temperature was maintained at 400 K using the v-rescale thermostat in the NVT ensemble. Electrostatic interactions were computed using the PME method. The FFT grid spacing was 0.1 nm. A cutoff distance of 1.2 nm was used in the calculation of electrostatic interactions in real space.

Each system was first equilibrated for 20 ns using a constant-charge simulation and then for another 5 ns using a constant potential simulation at zero cell voltage. To study the capacitive behavior, simulations were performed under cell voltages ranging from 0 to 5 V. To obtain charging dynamics, the cell voltage was set to 4 V in all cases and the systems were simulated for 40 ns to approach equilibrium as close as possible. We carried out more than three independent simulations of each system.

All simulations were performed using a customized version of the software GROMACS.[31] We used the constant potential method (CPM) to maintain a constant potential difference between the two electrodes during the simulation, which is important for charging dynamics.[32-33] We implemented the CPM in GROMACS following the methodology proposed by Siepmann *et al.* [34] and refined by Reed *et al.*[35]. In our CPM, the potential is applied directly on the electrode atoms, rather than on the electrode surface. This approach agrees well with the other implementations of the CPM. [36-37] In all simulations, the electrode charge was updated each time step.

We computed 1D diffusion coefficients from mean-square displacements (MSD)

$$D = \frac{1}{2t} < |z(t) - z_i(0)|^2 >, \qquad (1)$$

where $z_i(t)$ is the position of ion $i$ at time $t$ along z-axis, and $z_i(0)$ is its position at $t = 0$. Note that we calculated the MSDs for in-pore ions along the pore length for systems in equilibrium.

## Results and discussion

### Asymmetric characteristics of cathode-anode capacitance

We first consider supercapacitors with two identical single-pore electrodes, corresponding to single-peak pore-size distributions of porous electrodes. (**Fig. 1a**). To scrutinize the charge storage behavior, we calculated the integral capacitance, $C$, defined as: $C = Q/U$, where $Q$ is the surface charge density and $U$ is the potential of the electrode relative to the potential of zero charge (PZC, **Table S1**). **Figure 2a** shows that the capacitive behaviors of the $d_{0.50}//d_{0.50}$ and $d_{0.75}//d_{0.75}$ systems differ qualitatively. The $d_{0.50}//d_{0.50}$ system exhibits an asymmetric behavior: the capacitance under negative polarization is generally larger than at the positive. In sharp contrast, the capacitance of the $d_{0.75}//d_{0.75}$ system is symmetric with respect to the PZC and varies only slightly with the electrode potential. These differences can be understood by analyzing the number density of cations and anions at different electrode potentials (**Fig. 2b**). For the $d_{0.50}//d_{0.50}$ system, the charge storage is dominated by anion desorption (adsorption) under negative (positive) polarization. The charge storage in the $d_{0.75}//d_{0.75}$ system is driven by counterion adsorption and co-ion exchange under both negative and positive polarizations.

To quantify these features, we calculated the charging-mechanism parameter, $X$, defined as: [22]

$$X = \frac{N - N_0}{(N^{counter} - N^{co}) - (N_0^{counter} - N_0^{co})} \qquad (2)$$

where $N$ and $N_0$ are the total numbers of ions inside the pores at a non-zero electrode potential and at the PZC, respectively; similarly for counterions ($N^{counter}$ and $N_0^{counter}$) and co-ions ($N^{co}$ and $N_0^{co}$). $X$ equals to +1 (-1) for pure counterion absorption (co-ion desorption) and 0 for an exact one-to-one cation-anion exchange.

By trisecting the overall range of $X$ between -1 and +1 (note, however, that $X$ can also be smaller than -1 or larger than 1[38]), we further classified the charging mechanisms into the regions dominated by counterion adsorption, co-ion desorption and ion exchange. For $X > 1/3$, the change in the number of counterions is at least twice as large as the change in the number of co-ions, meaning that charging is dominated by counter-ion adsorption; similarly, the $X < -1/3$ region denotes the co-ion desorption domination. In the ion-exchange region (-1/3 < $X$ < 1/3), the change in the number of co-ions and counterions is comparable.

As shown in **Fig. 2c**, for system $d_{0.50}$//$d_{0.50}$, the $X$ values are mainly less than -1/3 under negative polarization and larger than 1/3 under positive polarization, indicating that the charge storage is driven by co-ion desorption and by counterion absorption, respectively. For system $d_{0.75}$//$d_{0.75}$, we found that -1/3 < $X$ < 1/3, suggesting that the charge storage is dominated by ion-exchange under both negative and positive polarization. Comparing $C$ and $X$ in **Fig. 2a** and **2c** for the $d_{0.50}$//$d_{0.50}$ system shows that the capacitance is higher when charge storage is driven by co-ion desorption (at negative potentials), compared to the capacitance when charge storage is due to counterion adsorption (at positive potentials). This is because the counterions from the bulk electrolyte must overcome an entropic barrier and unfavorable counterion-counterion interactions to enter a pore, while these barriers are weaker or absent in the case of co-ion desorption.[39] Interestingly, for the $d_{0.75}$//$d_{0.75}$ system, $X$ varies little with the electrode potential and the capacitance is nearly constant.

**Charging mechanisms of electrochemical cells**

To characterize the charging of a whole electrochemical cell, we calculated the net charging mechanism parameter, $\Delta X$,[40]

$$\Delta X = X_{pos} - X_{neg} \qquad (3)$$

where $X_{pos}$ and $X_{neg}$ are the charging-mechanism parameters of the positive and negative electrodes, respectively. Similarly to the categorization of $X$, we divided $\Delta X$ into three domains: cation domination ($\Delta X < -2/3$) and anion domination ($\Delta X > 2/3$), and the region $-2/3 < \Delta X < 2/3$ where the contributions from cation and anion rearrangements are comparable. The change in the number of in-pore cations and anions is similar in the region $-2/3 < \Delta X < 2/3$ and hence charging can be considered as symmetric; correspondingly, the regions $\Delta X < -2/3$ and $\Delta X > 2/3$ can be associated with asymmetric charging (**Fig. 2d**). For instance, for 1-methyl-1-propylpyrrolidinium bis(trifluoromethanesulfonyl)imide [Pyr$_{13}$][TFSI] IL in porous carbon electrodes, $X_{pos} = 0.3$ and $X_{neg} = 0.4$, as estimated from NMR measurements.[22] This yields $\Delta X = 0.7$ and implies asymmetric charging with the dynamics mainly driven by the motion of TFSI$^-$ ions.

**Figure 2d** shows that $\Delta X$ of system $d_{0.50}//d_{0.50}$ is over $2/3$, resulting in an asymmetric charge-storage performance, while $\Delta X$ of $d_{0.75}//d_{0.75}$ is close to zero, leading to the symmetric capacitance (**Fig. 2a**). The dissimilar response of EMIM$^+$ and BF$_4^-$ in the $d_{0.50}//d_{0.50}$ and $d_{0.75}//d_{0.75}$ systems likely originates from different nanoconfinement effects. In the 0.5 nm pore, there is only one layer of either cations or anions, and the ethyl group of EMIM$^+$ is closer to the pore wall than BF$_4^-$ (**Fig. S4a**), suggesting that EMIM$^+$ is more strongly confined than BF$_4^-$, and thus BF$_4^-$ could move more freely than EMIM$^+$. For a larger pore (0.75 nm pore in **Fig. S4b**), the pore has enough space to accommodate more than one layer of ions, so that the nanoconfinement effects on cations and anions become weaker. This is manifested by the cation mobility in these pores[41-42]:: at the PZC, the diffusion coefficient of EMIM$^+$ in the 0.75 nm pore is $12.6 \times 10^{-11}$ m$^2$ s$^{-1}$, which is more than 4 times larger than in the 0.5 nm pore ($3.0 \times 10^{-11}$ m$^2$ s$^{-1}$).

For the ($d_{0.45}//d_{0.45}$) system with ionophobic 0.45 nm pores, we found $X_{pos} = X_{pos} = 1$ and hence $\Delta X = 0$. Correspondingly, the capacitances of the negative and positive electrodes practically coincide (**Fig. S5**).

We also studied how asymmetry of electrodes affects the charge storage by analyzing simulation systems consisting of two electrodes with differently-sized pore (*i.e.*, asymmetric electrode systems). The capacitance of the $d_{0.50}//d_{0.75}$ system shows an asymmetric behavior (**Fig. 2e**), which can be associated with $\Delta X$ differing significantly from zero (**Fig. S6**). It is interesting to note that the capacitance of the $d_{0.75}//d_{0.50}$ system is more symmetric than that of the $d_{0.50}//d_{0.75}$ system (red circles in **Fig. S7** *vs.* **Fig. 2e**) and its $\Delta X$ is closer to zero (**Fig. S6**). Thus, the relation between the capacitance and $\Delta X$ for symmetric and asymmetric electrode systems suggests that $\Delta X$ can serve as an indicator for evaluating the asymmetric behavior of the capacitance: the symmetric ion response in the cathode and anode ($\Delta X \approx 0$) leads to the symmetric capacitance behavior, no matter whether the cathode and anode are identical or not. It is worth noting that the symmetric capacitance behavior can help to increase the operating voltage of supercapacitors.[16]

**"Equivalent principles" for supercapacitors**

Since in a supercapacitor the cathode and anode are separated by an electrolyte-filled separator,[43] resulting in a wide region with bulk electrolyte, the capacitances of an electrode in asymmetric and symmetric supercapacitors are equal under the same electrode potentials (**Fig. S8**). In this case, the capacitance of an overall system can be calculated from the capacitances of single electrodes using the equivalent capacitance. We demonstrate this for the $d_{0.50}//d_{0.75}$ system using the capacitances calculated from the $d_{0.50}//d_{0.50}$ and $d_{0.75}//d_{0.75}$ systems. The results show an excellent agreement, as expected (**Fig. 2e**). We obtained a similarly good agreement also for the $d_{0.75}//d_{0.50}$ system (**Fig. S7**).

Can a similar approach be applied to charging dynamics? To answer this question, we applied the popular transmission line model[44] (TLM, **Fig. S9**), frequently used to describe supercapacitor charging. We calculated the resistivities due to ionic transport in nanopore electrodes for the same symmetric and asymmetric supercapacitors as above. Unlike the equivalent capacitance, the ionic resistivities show significant deviations for both $d_{0.50}//d_{0.75}$ (**Fig. 2f**) and $d_{0.75}//d_{0.50}$ (**Fig. S10**) systems. This suggests that the charging dynamics of one electrode is affected by the dynamics processes occurring in the other electrode, which is evidenced by the time-evolution of the electrode potential that differs significantly for the symmetric and asymmetric supercapacitors (**Fig. S11**).

**Charging dynamics of supercapacitors with single-pore electrodes**

**Figure 3a** shows charging curves for supercapacitors with single-pore electrodes. To characterize the rate of charging, we estimated charging time $\tau$ as a time at which the amount of the accumulated charge became larger than 95% of the equilibrium value. As shown in **Fig. 3b**, the symmetric $d_{0.75}//d_{0.75}$ electrode system shows the fastest charging, with the charging time about 0.84 ns, while the charging of the $d_{0.50}//d_{0.50}$ system is the slowest (10.38 ns). The charging of the $d_{0.75}//d_{0.50}$ supercapacitor is several times faster than of the $d_{0.50}//d_{0.75}$ one (1.53 ns *vs.* 8.13 ns), demonstrating that the direction of the applied cell voltage has a considerable impact on the charging dynamics of the asymmetric electrode system.

To understand this charging behavior, we analyzed time-evolution of the ion number densities inside the pores. Charging dynamics of the $d_{0.50}//d_{0.50}$ system is ruled by co-ion desorption in the negative electrode and by counterion absorption in the positive electrode, and requires a long time (>20 ns) to reach equilibrium (**Fig. 3c**). For system $d_{0.75}//d_{0.75}$, the charging dynamics is dominated by ion exchange and the system reaches the equilibrium very quickly (<5 ns, **Fig. 3d**). For the asymmetric $d_{0.50}//d_{0.75}$ system, the charging of the positively charged 0.75 nm pore is slowed down by the negative 0.50 nm pore (**Fig. 3e**), while for the $d_{0.75}//d_{0.50}$ system, the charging of the 0.50 nm pore is sped up by the 0.75 nm pore (**Fig. 3f**). This also means that the 0.5 nm pore charges faster under positive polarization than under negative.

**In-pore ion diffusion**

To gain insights into the charging dynamics, we investigated in-pore ion diffusion under different polarizations. We first note that the 0.5 nm pore has a higher ion density when a positive potential is applied to the pore with respect to the PZC (**Fig. 3g**). Both modeling and experiments have demonstrated that a higher ion density results in slower ion diffusion.[26-27, 45] We observe similar trends also in our systems (**Fig. 3h**). We found that the cation diffusion coefficient in the 0.5 nm pore under negative polarization is comparable to the bulk diffusion coefficient ($51.5 \times 10^{-11}$ m$^2$ s$^{-1}$ vs. $53.8 \times 10^{-11}$ m$^2$ s$^{-1}$) and is much larger than at the PZC ($3.0 \times 10^{-11}$ m$^2$ s$^{-1}$). The cation diffusion coefficient under positive polarization $D=0.4 \times 10^{-11}$ m$^2$ s$^{-1}$ is much smaller than at the PZC. Similar trends are also observed for the anion diffusion (**Fig. 3h**).

Numerous studies have demonstrated that faster diffusing ions help accelerate charging dynamics in nanoporous electrodes.[30, 46-47] Perhaps surprisingly, our results demonstrate an opposite relation: in the 0.5 nm pore, charging is faster under negative polarizations when the pore is more occupied with ions and hence ion diffusion is slower.

**Why do the pores providing fast ion diffusion charge slower?**

To answer this question, we investigated time-evolution of the in-pore ion densities along the direction of the pore length (**Fig. 4** and **Fig. S14**). For the negatively charged 0.5 nm pore, there are more counterions than co-ions at the pore entrance (**Fig. 4b, c** and **Fig. S14a, b**). This is because initially, during the first 2 ns, while the co-ions diffuse out of the pore, the counter-ions enter the pore from the bulk electrolyte in a larger amount, mainly accumulating at the pore entrance. Thus, during this initial stage, overfilling occurs (**Fig. 3g**) that leads to co-ion trapping inside the pore. The co-ions have to get through this block of counter-ions at the pore entrance that causes their longer, jiggling motion, resulting in slow charging.[48]

Charging dynamics of the positively charged 0.5 nm pore is mainly driven by counter-ion adsorption, while co-ion desorption plays a minor role (**Fig. 3f**). The counterion adsorption and co-ion desorption are thus more balanced that helps avoid co-ion trapping (**Fig. 4e, f** and **Fig. S14c, d**). The counter-ions move towards the pore center, successively occupying the place of their fellow counterions that already moved more inside the pore (**Movie 2**). This process is similar to multi-ion concerted migration in ionic conductors, which can reduce energy barriers for ion motion, leading to higher ionic conductivity.[49-51] This explains why the 0.5 nm pore charges faster under negative polarizations, even though ion diffusion is faster at positive electrode potentials.

**Charging electrodes with multiple pores**

In practice, porous electrodes contain pores of different sizes.[52] To understand how the capacitance and charging dynamics of a pore are affected by the presence of other pores in the same electrode, we studied the charging of symmetric model supercapacitors having two electrodes, each consisting of two pores of different sizes (**Fig. 1b**). In this case, we found that the equivalent capacitance matches well with the MD results (**Fig. S16a, b**). This indicates that the capacitance of a complex electrode system with multiple pore sizes may be obtained by averaging the single pore results over the distribution of pore size.[53] Similarly as for the single-pore electrodes, the ionic resistivities computed within the TLM for the same pore in the single-pore and multiple-pore systems differ significantly (**Fig. S16c, d**), implying again that the charging dynamics cannot be easily predicted from the single-pore results.

We thus investigated the charging dynamics of electrodes with multiple pore sizes. **Figure 5a** demonstrates that the charging curve for the $d_{0.50}d_{0.75}//d_{0.50}d_{0.75}$ system is between the curves for the corresponding symmetric electrode systems with single pore sizes. We found the charging time (4.17 ns) is in-between the charging times for the single-pore systems (0.84 ns for $d_{0.75}//d_{0.75}$ and 10.38 ns for $d_{0.50}//d_{0.50}$). To understand the electrode charging better, we investigated the charging dynamics of each pore in the $d_{0.50}d_{0.75}//d_{0.50}d_{0.75}$ system.

**Figure 5b** shows that under both negative and positive polarizations, the 0.5 nm pore of the $d_{0.50}d_{0.75}//d_{0.50}d_{0.75}$ electrode charges faster than the same pore in the $d_{0.50}//d_{0.50}$ system, suggesting that slow charging of a single pore electrode can be accelerated in a system with multiple pore sizes. Charging of the 0.5 nm pore is faster under positive polarization, in agreement with the results for the asymmetric electrode systems ($d_{0.75}//d_{0.50}$ *vs.* $d_{0.50}//d_{0.75}$, see **Fig. 3a**). Under negative potentials, the total ion density shows a small overfilling in both systems (**Fig. 5c**). As discussed, overfilling causes clogging at the pore entrance, leading to sluggish dynamics. However, defilling of this pore occurs faster in the multi-pore electrode system than in the single-pore $d_{0.50}//d_{0.50}$ system, making its charging faster.

In contrast with the 0.5 nm pore, charging of the positively 0.75 nm pore is slowed down in the multi-pore electrodes, compared with the single-pore $d_{0.75}//d_{0.75}$ system (**Fig. 5d**). Under negative polarization, we observe overcharging, that is, initially the charge increases quickly to a value above equilibrium, and approaches the equilibrium value from above (**Fig. 5d**). We recall that we did not detect overcharging in single-pore electrodes. It is noteworthy that overcharging has been reported for nano-sized cells with flat electrodes due to double-layer overlaps[54] and for flat electrodes driven by non-local screening and degenerate mobility of concentrated ionic systems[55]. In our case, we note that the electrode becomes more polarized initially, which drives an excess amount of counterions into the 0.75 pore (**Fig. 5e**). The narrower 0.5 nm pore of the same electrode is more densely populated by ions and charges slower, which is likely the reason why overcharging does not occur in this pore.

Overcharging of the 0.75 nm pore is accompanied by counter-ion overfilling and co-ions over-defilling, that is, the density of co-ions decreases to a value below the equilibrium (**Fig. 5f**). However, these two processes are much faster than defilling of the 0.5 nm pore (**Fig. 5c**). This is because the 0.75 nm pore is less crowded, hence the adsorbed counter-ions can move quicker inside the pore (**Fig. S17**). In contrast, the crowded 0.5 nm pore produces congestion at the pore entrance that blocks the co-ions from diffusing out of the pore, thus slowing down the overall charging process.

We found that overfilling and overcharging are present in the $d_{0.45}d_{0.50}//d_{0.45}d_{0.50}$ supercapacitor (**Fig. S19**), but not in the $d_{0.45}d_{0.75}//d_{0.45}d_{0.75}$ one (**Fig. S20**). This is likely because in the latter case the net charging mechanism parameter $\Delta X$ vanishes, while it is essentially non-zero for the $d_{0.50}d_{0.75}//d_{0.50}d_{0.75}$ and $d_{0.45}d_{0.50}//d_{0.45}d_{0.50}$ systems ( $\Delta X = 0.67$ and $\Delta X = 0.59$, respectively). This observation suggests a relation between the dynamics and charging mechanisms that we discuss next.

**How charging mechanisms are related to power density and energy storage**

Our analysis suggests a correlation between the net charging mechanism parameter $\Delta X$ and charging dynamics. In particular, supercapacitors characterized by small $\Delta X$ provide faster charging than systems with larger $\Delta X$. For instance, for single-pore electrodes, the $d_{0.45}//d_{0.45}$ system has $\Delta X = 0$ and the charging time $\tau_c = 0.79$ ns, while for the $d_{0.50}//d_{0.50}$ system $\Delta X = 1.03$ and $\tau_c = 10.38$ ns. Similarly, for multi-pore electrodes, the $d_{0.45}d_{0.75}//d_{0.45}d_{0.75}$ system has $\Delta X$ close to zero and $\tau_c = 0.79$ ns, while $\Delta X = 0.67$ and $\tau_c = 4.29$ ns for the $d_{0.50}d_{0.75}//d_{0.50}d_{0.75}$ systems. This is summarized in **Fig. 6a** for all systems considered in this work. The correlations are apparent and can be intuitively understood by associating charging mechanisms with charging dynamics. A vanishing net charging mechanism parameter means that the charging mechanisms in both electrodes are balanced and involve similar processes (for example, cation domination), likely characterized by similar dynamics. A non-zero $\Delta X$ implies different charging mechanisms between cathode and anode, whereby one electrode may slow down the other, thus slowing down the whole charging process.

We also estimated the energy density stored in a supercapacitor at cell voltage $U$,

$$E(U) = \int_0^U C_D(u)\, u\, du = UQ(U) - \int_0^U Q(u)\, du, \qquad (4)$$

where $C_D$ is the differential capacitance. To obtain $E(U)$, we used the MD data and performed the integration in **Eq (4)** numerically. We note, however, that due to the limited number of data points, our results shall only be treated as a crude estimation of the stored energy density. Nevertheless, these results suggest no relation between energy storage and asymmetry of the charging mechanisms between cathode and anode (**Fig. S21a**). This is because $\Delta X$ does not describe the actual charging mechanisms, but whether the charging mechanisms of the two electrodes are symmetrized.

From the stored energy density and charging times, we estimated the power density as $P = E/\tau_c$. Similarly to the charging times (**Fig. 6a**), we found that the power density correlates with the net charging mechanism parameter (**Fig. S21b**). **Figure 6b** shows the Ragone plot, which demonstrates that a symmetric supercapacitor based on the narrowest 0.45 nm pore, which is ionophobic, provides the highest energy and power densities. This is in accord with previous works[30, 40], showing that ionophobic pores can enhance both energy and power density. A supercapacitor with two identical electrodes having the widest pore (0.75 nm) has the lowest energy density and moderate power density. Combining such electrodes with narrower pores enhances energy storage but reduces the power density, demonstrating the power-density tradeoff of a supercapacitor. However, combining these pores with ionophobic pores within the same electrode (system $d_{0.45}d_{0.75}//d_{0.45}d_{0.75}$) again enhances both the energy and power densities.

## Conclusion

We have investigated asymmetric features of nanoporous supercapacitors and their effect on the capacitance and charging dynamics. We focused on the asymmetry of porous electrodes and ion response and found that the capacitance and charging dynamics correlate with the charging mechanisms at the cathode and anode, no matter whether the electrodes are identical or not. We showed that a small or vanishing net charging mechanism parameter ($\Delta X \approx 0$, **Eq (3)**) is associated with the symmetric capacitance behavior (**Fig. 2a, b**) and that $\Delta X$ correlates with the speed of charging (**Fig. 6a**), which may find practical applications in designing an optimal supercapacitor.

For supercapacitors with asymmetric electrodes, our simulation revealed that the direction of an applied cell voltage can drastically affect their charging dynamics and hence power density. Perhaps surprisingly, we found that a negatively charged narrow pore can charge slowly, despite fast in-pore ion diffusion (**Fig. 3**). We attributed this effect to overfilling, occurring during the initial stage of charging and leading to longer, jiggling motion of co-ions before they diffuse out of the pore. In contrast, for the positively charged pore, the overfilling is absent and the multi-ion concerted migration-type of motion causes shorter and straighter ion motion paths, resulting in faster charging dynamics. We found the overfilling also takes place in multi-pore electrodes, but the following defilling is faster than in the single-pore electrodes, which accelerates the charging dynamics. We also revealed that charging of multi-pore electrodes can be accompanied by overcharging, during which the accumulated charge grows over the equilibrium value (**Fig. 5d**).

We explored the relation between supercapacitor systems with single-pore electrodes and systems with asymmetric electrodes and electrodes with multiple pore sizes. We confirmed that the equivalent capacitance approach provides quantitatively correct results, allowing one to calculate the capacitance of electrodes with multiple pore sizes using the results for single-pore electrodes. However, the in-pore resistivities, as calculated by fitting the MD results to the transmission line model, did not show such simple relations, suggesting that there is no straightforward way to predict the charging dynamics of complex nanoporous electrodes (**Fig. 2f**).

All in all, our work built a bridge between understanding charging in single-sized pores and in nanoporous electrodes, which is vital for designing supercapacitors with high energy and power densities.

## Acknowledgments

Authors acknowledge the funding support from the National Natural Science Foundation of China (51876072) and the Hubei Provincial Natural Science Foundation of China (2019CFA002, 2020CFA093). The authors also thank Beijing PARATERA Tech CO., Ltd. for providing HPC resources to accomplish simulations in this work. This work is also supported by the Program for HUST Academic Frontier Youth Team.

## Supplementary Information

Details of MD simulation, additional results of ion structure inside pores, capacitance and charging dynamics, movies of charging process. This material is available free of charge *via* the Internet at http://...

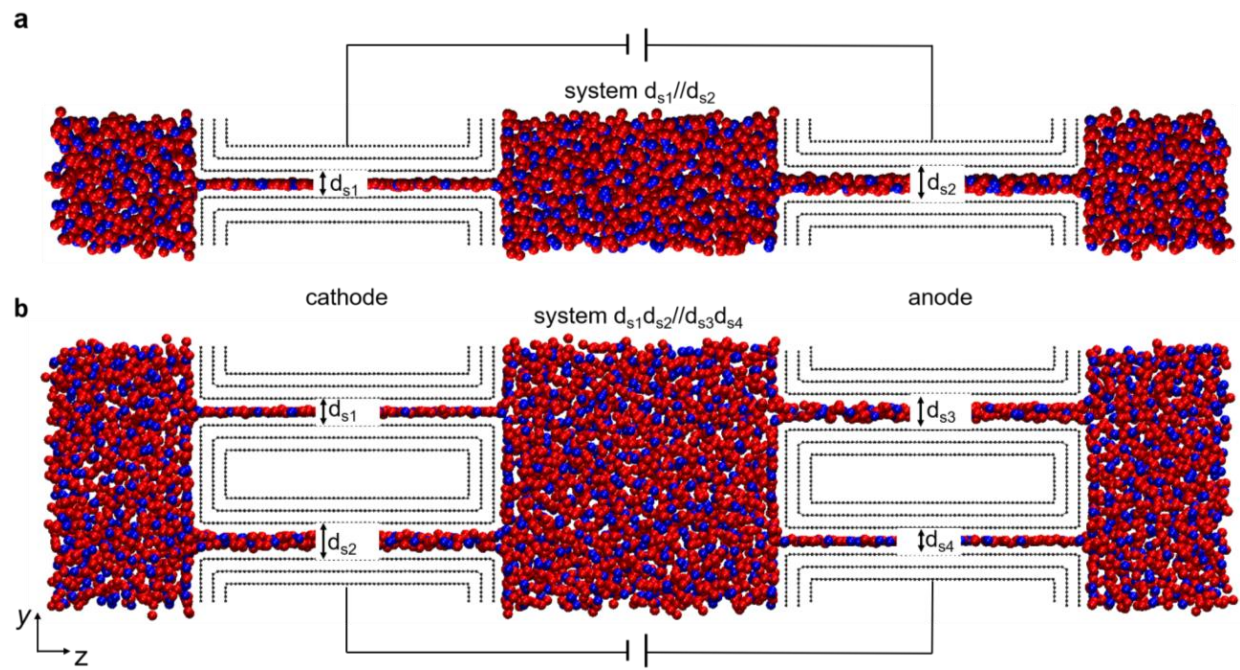

**Figure 1. Molecular dynamics simulation system setup. a**, Snapshot of a system with single pore size. The system is named $d_{s1}//d_{s2}$, where s1 and s2 are the pore sizes of the negative (cathode) and positive (anode) electrodes. **b,** Snapshot of a system with multiple pore sizes. The system is named $d_{s1}d_{s2}//d_{s3}d_{s4}$, specifying that the system consists of the negatively charged electrode (cathode) with pores of sizes s1 and s2, and of the positively charged electrode (anode) with pores of sizes s3 and s4 (unit: nm). For all systems, the left electrode is cathode and the right one is anode. The black spheres represent carbon atoms, and the colored spheres represent the coarse-grained model of [EMIM][BF$_4$].

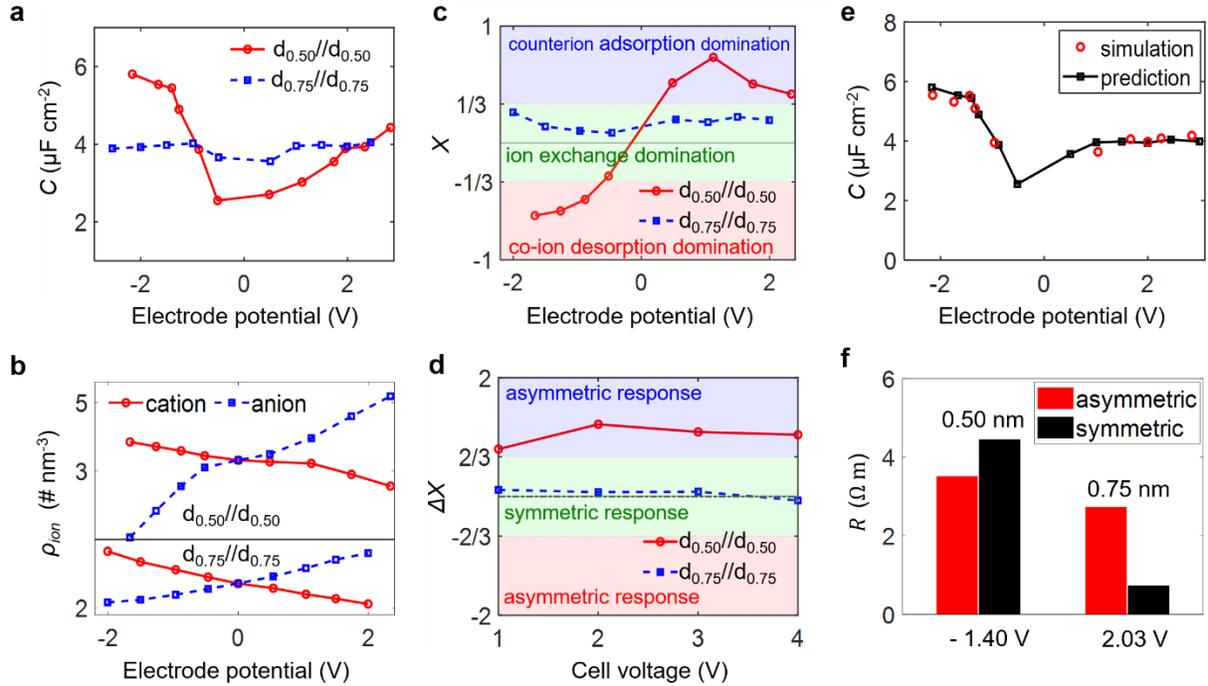

**Figure 2. Asymmetric behavior of capacitance. a,** Integral capacitance, $C$, of symmetric electrode systems versus the electrode potential. **b,** Number densities of cation and anion, $\rho_{ion}$, as functions of the electrode potential. **c,** Charging mechanism parameter, $X$, as a function of the electrode potential. **d,** Net charging mechanism parameter, $\Delta X$, as a function of the cell voltage. **e,** Integral capacitance, $C$, of asymmetric electrode system $d_{0.50}//d_{0.75}$ obtained by MD simulations and predicted by the equivalent capacitance from the results for the $d_{0.50}//d_{0.50}$ and $d_{0.75}//d_{0.75}$ systems. The positive (negative) electrode potentials correspond to the capacitance of the positive (negative) electrodes. **f,** Resistivities of ionic transport, $R$, of the negative (0.5 nm) and positive (0.75 nm) pores of the asymmetric $d_{0.50}//d_{0.75}$ system and of the symmetric $d_{0.50}//d_{0.50}$ and $d_{0.75}//d_{0.75}$ systems obtained by fitting the MD results to the transmission line model.

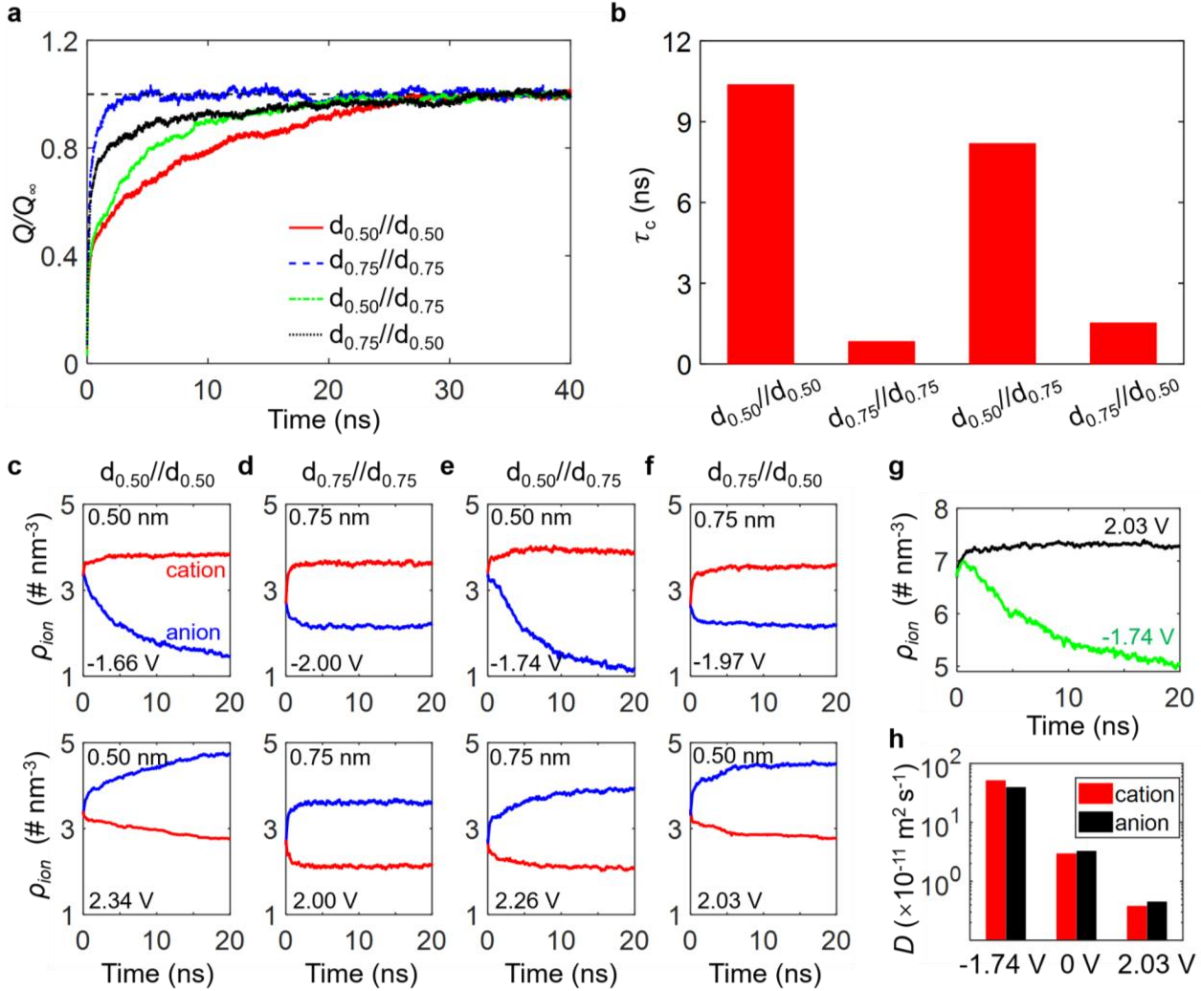

**Figure 3. Charging dynamics of systems with single pore size. a,** Comparison of charging process of different systems with a cell voltage of 4 V. $Q_\infty$ is the electrode surface charge density at equilibrium. **b,** Charging times, $\tau_c$, of different systems. **c-f,** Evolution of the number densities of cations and anions, $\rho_{ion}$, of systems $d_{0.50}//d_{0.50}$ (**c**), $d_{0.75}//d_{0.75}$ (**d**), $d_{0.50}//d_{0.75}$ (**e**) and $d_{0.75}//d_{0.50}$ (**f**) under negative (**upper plane**) and positive (**lower plane**) polarization. Red lines represent the number density of cation and blue lines represent the number density of anion. **g,** Evolution of the total ion number density, $\rho_{ion}$, of the negatively charged (-1.74 V) 0.50 nm pore of the $d_{0.50}//d_{0.75}$ system (green line) and the positively charged (2.03 V) 0.50 nm pore of the $d_{0.75}//d_{0.50}$ system (black line). **h,** Diffusion coefficient, $D$, of cations and anions inside the negatively charged (-1.74 V) 0.50 nm pore of the $d_{0.50}//d_{0.75}$ system, inside the 0.50 nm pore at PZC (0 V), and inside the positively charged (2.03 V) 0.50 nm pore of the $d_{0.75}//d_{0.50}$ system.

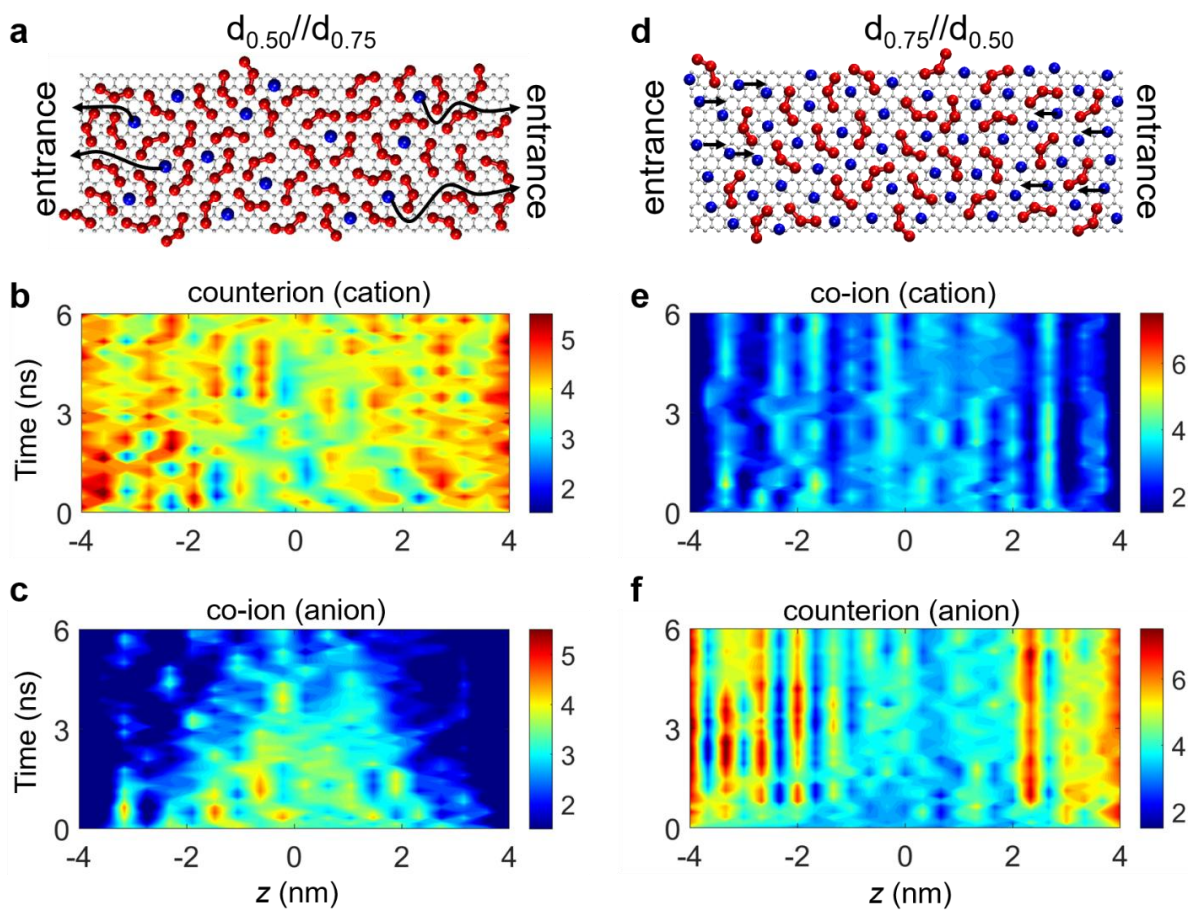

**Figure 4. Ion motion paths inside 0.5 nm pores. a,** Top-view snapshot of the negatively charged 0.50 nm pore of system $d_{0.50}//d_{0.75}$. **b-c,** Time-evolution of cations (**b**) and anions (**c**) inside the negatively charged 0.50 nm pore of the $d_{0.50}//d_{0.75}$ system along the direction of the pore length, $z$. **d,** Top-view snapshot of ions inside the positively charged 0.50 nm pore of the $d_{0.75}//d_{0.50}$ system. **e-f,** Time-evolution of cations (**e**) and anions (**f**) inside the positively charged 0.50 nm pore of the $d_{0.75}//d_{0.50}$ system, respectively. Red spheres and blue spheres represent cations and anions. Grey spheres are carbon atoms. Black arrows show schematically the motion paths of anions. $z = 0$ is the center of the pore along the direction of pore length. Unit of colorbar: #/nm$^3$.

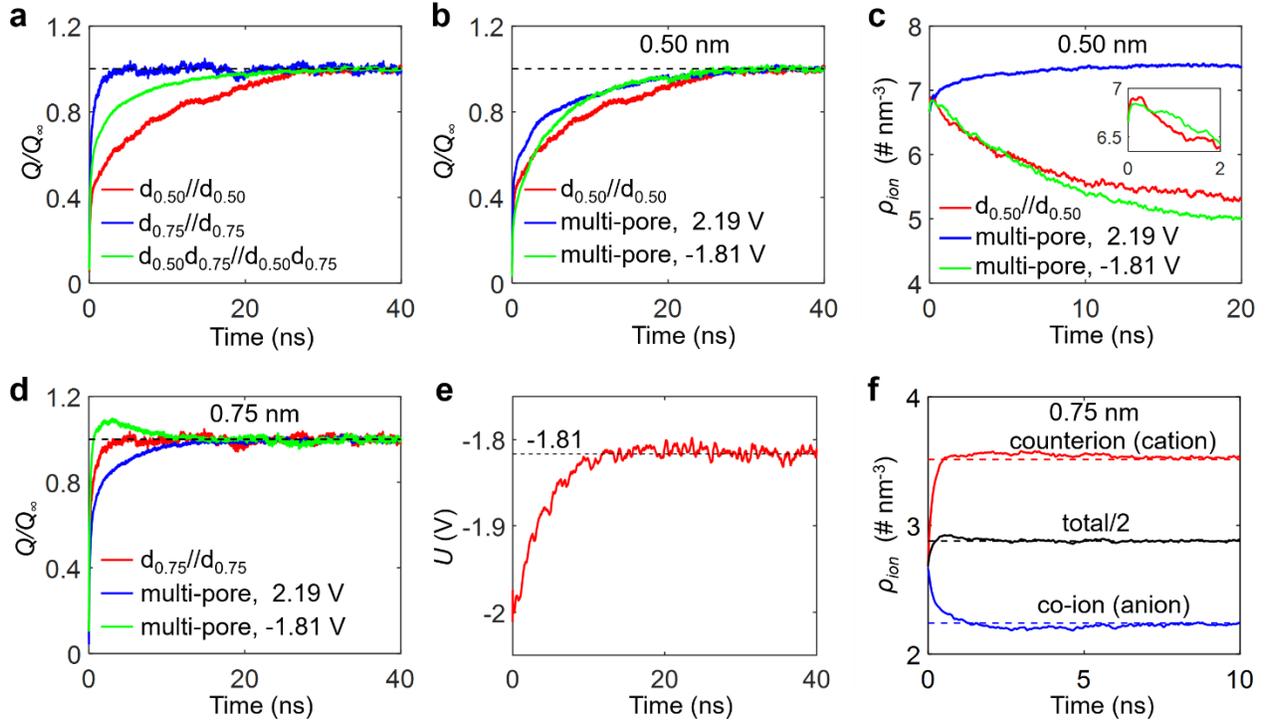

**Figure 5. Charging dynamics of systems with multiple pore sizes. a,** Comparison of the charging process of the $d_{0.50}d_{0.75}//d_{0.50}d_{0.75}$ system (multiple pore sizes) and systems with single pore size. **b,** Comparison of the charging process inside the 0.50 nm pore of the $d_{0.50}d_{0.75}//d_{0.50}d_{0.75}$ and $d_{0.50}//d_{0.50}$ systems. **c,** Comparison of the ion number density, $\rho_{ion}$, inside the 0.50 nm pore of system $d_{0.50}d_{0.75}//d_{0.50}d_{0.75}$ and the negatively charged 0.50 nm pore of system $d_{0.50}//d_{0.50}$. **d,** Comparison of the charging process inside the 0.75 nm pore of systems $d_{0.50}d_{0.75}//d_{0.50}d_{0.75}$ and $d_{0.75}//d_{0.75}$. **e,** Electrode potential, $U$, at the negative electrode of the $d_{0.50}d_{0.75}//d_{0.50}d_{0.75}$ system versus time. The red solid line shows the time-evolution of the electrode potential and the black dotted line is the electrode potential at equilibrium. **f,** Ion number density, $\rho_{ion}$, inside the negatively charged 0.75 nm pore of the $d_{0.50}d_{0.75}//d_{0.50}d_{0.75}$ system.

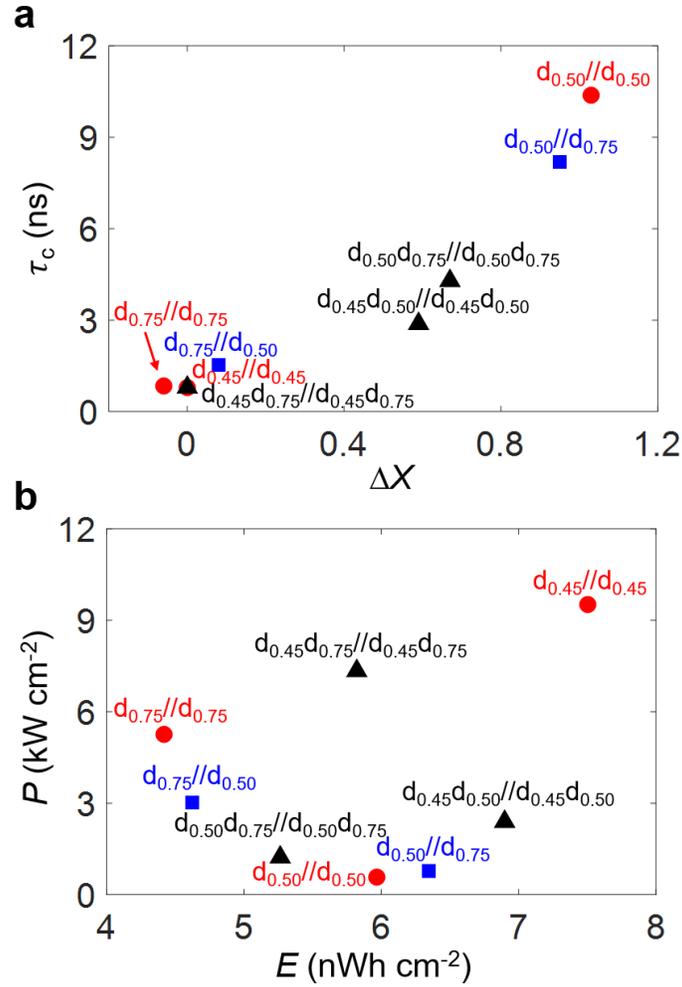

**Figure 6. Charging time, power density and energy storage. a,** Correlation between the charging time, $\tau_c$, and the net charging mechanism parameter, $\Delta X$. **b,** Power density, $P$, and energy density, $E$, of different systems. Red dots and blue squares show the data for the symmetric and asymmetric electrode systems with single-pore electrodes, respectively, and black triangles denote the data for the systems with multi-pore electrodes.

## Graphical abstract

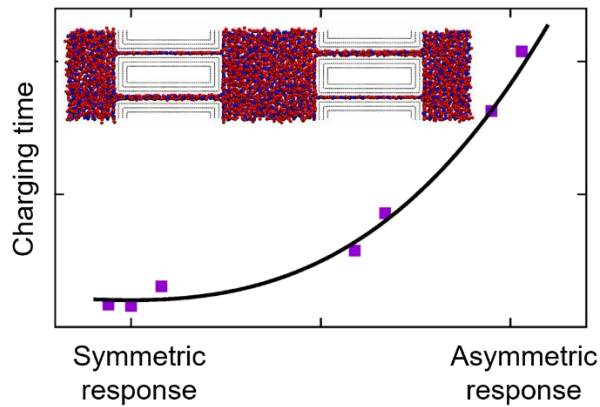

Constant-potential molecular dynamics simulations of symmetric and asymmetric nanoporous supercapacitors reveal that a symmetric ion response in the cathode and anode can boost power density.

## Highlights

- Capacitance and charging dynamics of supercapacitors correlate with ion response in the cathode and anode.

- Symmetrizing ion response in the cathode and anode can boost the charging dynamics of supercapacitors.

- Charging dynamics of narrow pores can be fast despite slow in-pore ion diffusion.

- Over-charging occurs in electrodes with multiple pore sizes, which speeds up their charging dynamics.